\documentclass[traditabstract]{aa}
\usepackage{graphicx}
\usepackage{txfonts}
\usepackage{epsfig}
\usepackage{natbib}
\usepackage{color}

\begin{document}

\title{Radial migration does little for Galactic disc thickening}

\author{I.~Minchev\inst{1}, 
B.~Famaey\inst{2,3}, 
A.~C.~Quillen\inst{4},
W.~Dehnen\inst{5},
M.~Martig\inst{6},
A.~Siebert\inst{2}
}

\institute{Leibniz-Institut f\"{ur} Astrophysik Potsdam (AIP), An der Sternwarte 16, D-14482, Potsdam, Germany
\email{iminchev1@gmail.com}
\and
Universit\'e de Strasbourg, CNRS, Observatoire Astronomique, 11 rue de l'Universit\'e, 67000 Strasbourg, France
\and
AIfA, University of Bonn, Germany
\and
Department of Physics and Astronomy, University of Rochester, Rochester, NY 14627
\and
Department of Physics \& Astronomy, University of Leicester, Leicester LE1 7RH, UK
\and
Centre for Astrophysics \& Supercomputing, Swinburne University of Technology, P.O. Box 218, Hawthorn, VIC 3122, Australia}

\date{Accepted 15 October 2012}

\abstract{
Non-axisymmetric components, such as spirals and central bars, play a major role in shaping galactic discs. An important aspect of the disc secular evolution driven by these perturbers is the radial migration of stars. It has been suggested recently that migration can populate a thick-disc component from inner-disc stars with high vertical energies. Since this has never been demonstrated in simulations, we study in detail the effect of radial migration on the disc velocity dispersion and disc thickness, by separating simulated stars into migrators and non-migrators. We apply this method to three isolated barred Tree-SPH N-body galaxies with strong radial migration. Contrary to expectations, we find that as stellar samples migrate, on the average, their velocity dispersion change (by as much as 50\%) in such a way as to approximately match the non-migrating population at the radius at which they arrive. We show that, in fact, migrators suppress heating in parts of the disc. To confirm the validity of our findings, we also apply our technique to three cosmological re-simulations, which use a completely different simulation scheme and, remarkably, find very similar results. We believe the inability of migration to thicken discs is a fundamental property of internal disc evolution, irrespective of the migration mechanism at work. We explain this with the approximate conservation of the (average) vertical and radial actions rather than the energy. This ``action mixing" can be used to constrain the migration rate in the Milky Way: estimates of the average vertical action in observations for different populations of stars should reveal flattening with radius for older groups of stars. 
}

\titlerunning{Migration and thick discs}
\authorrunning{I. Minchev et al.}

\maketitle

\section{Introduction}
\label{sec:intro}

Most local galaxies are observed to host a thick disc component (e.g., \citealt{comeron11}), including our Milky Way (MW), as first proposed by \cite{gilmore83}. Compared the thin disc, the thick disc is composed of older, metal-poor, alpha enriched stars. An important question of modern galactic astronomy is how thick discs came into existence.  

One possibility is that they were born thick at high redshift from the internal gravitational instabilities in gas-rich, turbulent, clumpy discs \citep{bournaud09} or in the turbulent phase associated with numerous gas-rich mergers \citep{brook04,brook05}. They could also have been created through accretion of galaxy satellites \citep{meza05,abadi03}, where thick disc stars then have an extragalactic origin. 

Another possibility is that thick discs are created through the heating of preexisting thin discs. This can either happen fast, on a Gyr timescale, in an early violent epoch, or as a continuous process throughout the galaxy lifetime. In the first case, the thick disc would appear as a clear distinct component in chemistry and phase-space, while in the latter case it would rather be seen as a gradual, continuous transition. Observationally, whether the MW thick disc is consistent with the former or the latter is heavily debated (e.g, \citealt{schonrich09, bovy12a}).  

Preexisting thin discs can be heated fast through the effect of multiple minor mergers \citep{quinn93, villalobos08, dimatteo11}, the rate of which decreases with decreasing redshift. Evidence for merger encounters can be seen in structure in phase-space of MW disc stars (e.g., \citealt{minchev09, gomez12a, purcell11}), which could last for as long as 4~Gyr \citep{gomez12b}. The recent work by \cite{minchev12b} presented numerical results indicating that the velocity dispersion of the oldest MW-disc stars cannot be obtained without substantial merger activity at high redshift.

The fact that galactic discs heat with time has now been known for over 60 years, established from both observational and theoretical work. To explain the observed correlation between the ages and velocity dispersions of solar neighborhood stars, \cite{spitzer51, spitzer53} suggested that massive gas clouds (then undetected) scattered stars from initially circular, into more eccentric and inclined, orbits. Giant molecular clouds were thought to be the sole scattering agents (e.g., \citealt{mihalas81}) until \citet{lacey84} showed that the observed ratio of the dispersion in the direction perpendicular to the Galactic plane and that toward the Galactic center, ${\sigma_z / \sigma_R}$, was too low to be consistent with the predictions from this scattering process. This resulted in the development of models that incorporated the heating of the stellar disc from transient spiral structure \citep{barbanis67,carlberg84} in addition to scattering from molecular clouds \citep{jenkins90,jenkins92}. Other proposed models for the heating of stars include scattering by halo black holes \citep{lacey85} or dark clusters \citep{carr87}, scattering by giant molecular clouds and halo black holes \citep{hann02, hann04}, and vertical disc heating through ``popping" star clusters \citep{kroupa02, kroupa05} or through ``levitation" caused by the 2:2 resonance between the horizontal and vertical oscillations of stars in an inside-out forming disc, converting their radial actions into vertical actions \citep{sridhar96a,sridhar96b}. An investigation of the origin of the age-velocity relation in the MW was presented recently by \cite{hause11}, by comparing with cosmological simulations. Interestingly, a recent work has shown that the age-velocity relation could arise not because of external perturbers or the gradual heating of a thin disc, but because the velocity dispersion of newly born stars decreases with redshift \citep{forbes11}. Finally, a viable internal mechanism for disc heating is the interaction of multiple spiral density waves, as proposed by \cite{mq06} or bar and spirals \citep{mf10}. Multiple waves are now known to exist in both observation \citep{elmegreen92, rix95, meidt09} and simulations (e.g., \citealt{masset97, quillen11, minchev12a, grand12, roskar11}), therefore this heating process may be ubiquitous in disc galaxies.

Interestingly, non-axisymmetric patterns also give rise to another important process regarding the secular evolution of galaxy discs: radial migration. This process is related to the redistribution of angular momentum of stars within the disc, generally resulting in an increase in disc scale-length. Several radial migration mechanisms have been described in the literature: (i) the effect of the corotation resonance of transient spiral density waves \citep{sellwood02, roskar11}, (ii) the effect of the non-linear coupling between multiple spiral waves \citep{mq06} or bar and spirals \citep{mf10,minchev11a, brunetti11}, and (iii) perturbations due to minor mergers \citep{quillen09, bird12}. Note that in case (ii) above, spirals do not need to be short-lived transients for efficient migration; relatively long-lived waves are expected to exist in barred discs \citep{bt08, quillen11,minchev12a}, but are also found in non-barred simulations (e.g. \citealt{donghia12}). Recently, \cite{comparetta12} showed that, even if patterns are long-lived, radial migration can result from short-lived density peaks arising from interference among density waves overlapping in radius. This ``disc mixing" can give rise to a number of phenomena, such as flattening in radial metallicity gradients (e.g., \citealp{minchev11a, pilkington12}) and extended stellar density profiles (e.g., \citealp{roskar08b, sanchez09, minchev12a}).  

Recently there has been a growing conviction that radial migration can result in a thick disc formation by bringing out high-velocity-dispersion stellar populations from the inner disc and the bulge. Such a scenario was used, for example, in the analytical model of \cite{schonrich09}, where the authors managed to explain the Milky Way thick- and thin-disc characteristics (both chemical and kinematical) without the need of mergers or any discrete heating processes. Similarly, the increase of disc thickness with time found in the simulation by \cite{roskar08a} has been attributed to migration in the works by \cite{sales09} and \cite{loebman11}. We have to point out, however, that how exactly radial migration affects disc thickening in dynamical models has never been demonstrated. 

In this paper we would like to examine in detail the effect of radial migration on the increase of radial and vertical velocity dispersion and, thus, on its possible relation to the formation of thick discs. For this purpose we study three isolated galaxy simulations and three simulations in the cosmological context. We must note that, while the simulations we examine address some of the most widely used initial conditions for simulating galactic discs: preassembled N-body discs and discs growing in the cosmological context, in this paper we are not testing the validity of other models of formation, such as the one used by \cite{sales09} and \cite{loebman11}.

\begin{figure*}
\includegraphics[width=18cm]{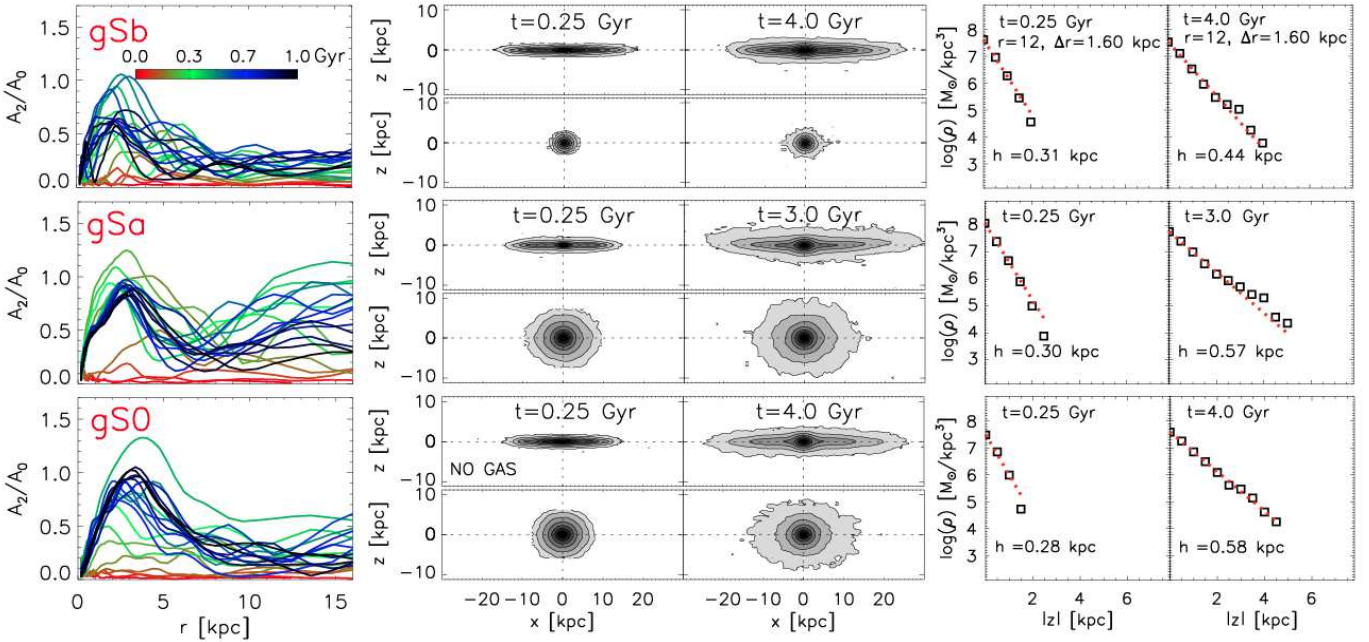}
\caption{
{\bf First column:} m=2 Fourier amplitudes, $A_2$, as a function of time (color bar) for our model galaxies. {\bf Columns 2-3:} edge-on view for each disc and bulge (separately) for earlier and later times of their evolution, as indicated in each panel. {\bf Columns 4-5:} scale-heights, $h$, at $\sim2.5$ disc scale-lengths. The dotted red line shows a single exponential fit. Note that in 3-4~Gyr of evolution with strong perturbation from bars and spirals, $h$ only doubles despite the efficient stellar radial migration.
}
\label{fig:1}
\end{figure*}

\section{Description of the simulations}
\label{sec:sims}

\subsection{Tree-SPH N-body simulations}
\label{sec:nbody}

For the majority of this paper we study three main runs of isolated disc galaxies from the GalMer database \citep{dimatteo07}: the giant S0, Sa, and Sbc runs (hereafter gS0, gSa, and gSb). In GalMer, for each galaxy type, the initial halo and the optional bulge are modeled as Plummer spheres, with characteristic masses $M_H$ and $M_B$, and characteristic radii $r_H$ and $r_B$, respectively. Their densities are given by
\begin{equation}\label{halo}
\rho_{H,B}(r)=\left(\frac{3M_{H,B}}{4\pi r^3_{H,B}}\right)\left(1+\frac{r^2}{r^2_{H,B}}\right)^{-5/2}.
\end{equation}

On the other hand, the initial gaseous and stellar discs follow Miyamoto-Nagai density profiles with masses $M_{g}$ and  $M_s$, and vertical and radial scale-lengths given by $b_{g}$ and $a_{g}$, and $b_s$ and $a_s$, respectively:
\begin{equation}
\begin{array}{l}
\rho_{g,s}(R,z)=\left(\frac{b^2_{g,s} M_{g,s}}{4 \pi}\right) \times \nonumber\\
\frac{a_{g,s} R^2+\left(a_{g,s}+3\sqrt{z^2+b^2_{g,s}}\right)\left(a_{g,s}+\sqrt{z^2+b^2_{g,s}}\right)^2 }
{ \left[a_{g,s}^2+\left(a_{g,s}+\sqrt{z^2+b^2_{g,s}}\right)^2\right]^{5/2}\left(z^2+b^2_{g,s} \right)^{3/2} }.
\end{array}
\end{equation}

The parameters for the initial conditions of the three runs, including the number of particles used for the gaseous disc, $N_g$, the stellar disc and bulge, $N_s$, and the dark matter halo, $N_{DM}$, are given in Table~\ref{parameters}. The initial Toomre parameter of both stars and gas is taken to be $Q=1.2$ as the initial condition of the Tree-SPH simulations and particle velocities are initialized with the method described in \cite{hern93}. Full details of the simulations are given in \cite{dimatteo07} and \cite{chilingarian10}.

  \begin{table}
      \caption[]{Galaxy modeling parameters}
         \label{parameters}
         \centering
         \begin{tabular}{lccc}
            \hline\hline
            & gS0 &gSa & gSb\\
            \hline
            $M_{B}\ [2.3\times 10^9 M_{\odot}]$ & 10 & 10 & 5\\
            $M_{H}\ [2.3\times 10^9 M_{\odot}]$ & 50 & 50 & 75\\
            $M_{s}\  [2.3\times 10^9 M_{\odot}]$ & 40 & 40 & 20\\
            $M_{g}/M_{s}$ & --& 0.1 & 0.2\\
            $r_{B}\ [\mathrm{kpc}]$ & 2 & 2 & 1\\
            $r_{H}\ [\mathrm{kpc}]$ & 10 & 10 & 12\\
            $a_{s }\ [\mathrm{kpc}]$  & 4 & 4 & 5\\
            $b_{s }\ [\mathrm{kpc}]$  & 0.5 & 0.5 & 0.5\\
            $a_{g }\ [\mathrm{kpc}]$  & -- & 5 & 6\\
            $b_{g }\ [\mathrm{kpc}]$ & -- & 0.2 & 0.2\\
             \hline
            $N_g$  & --          & 80\,000   & 160\,000\\
	    $N_{s}$ & 320\,000 & 240\,000 & 160\,000\\
	    $N_{DM}$ & 160\,000 & 160\,000 & 160\,000\\
            \hline
         \end{tabular}
   \end{table}

These galaxies have been studied extensively in \cite{minchev11a,minchev12a}. In these Tree-SPH N-body simulations the gSb model develops a bar similar to the one seen in the Milky Way (MW) \citep{mnq07,minchev10}, while both the gS0 and gSa bars are substantially larger than that (see Fig.~\ref{fig:1}). We have shown \citep{minchev11a} that this has strong effects on the migration efficiency found in these discs, with larger bars giving rise to stronger mixing (also consistent with studies of diffusion coefficients, e.g., \citealt{brunetti11,shevchenko11}). Most recently, we \citep{minchev12a} demonstrated that, in these same models, the strongest changes of angular momentum occur near the bar's corotation (CR) region, in contradiction to the expectation that patterns need to be transient for efficient migration \citep{sellwood02}.  

\begin{figure*}
\centering
\includegraphics[width=15cm]{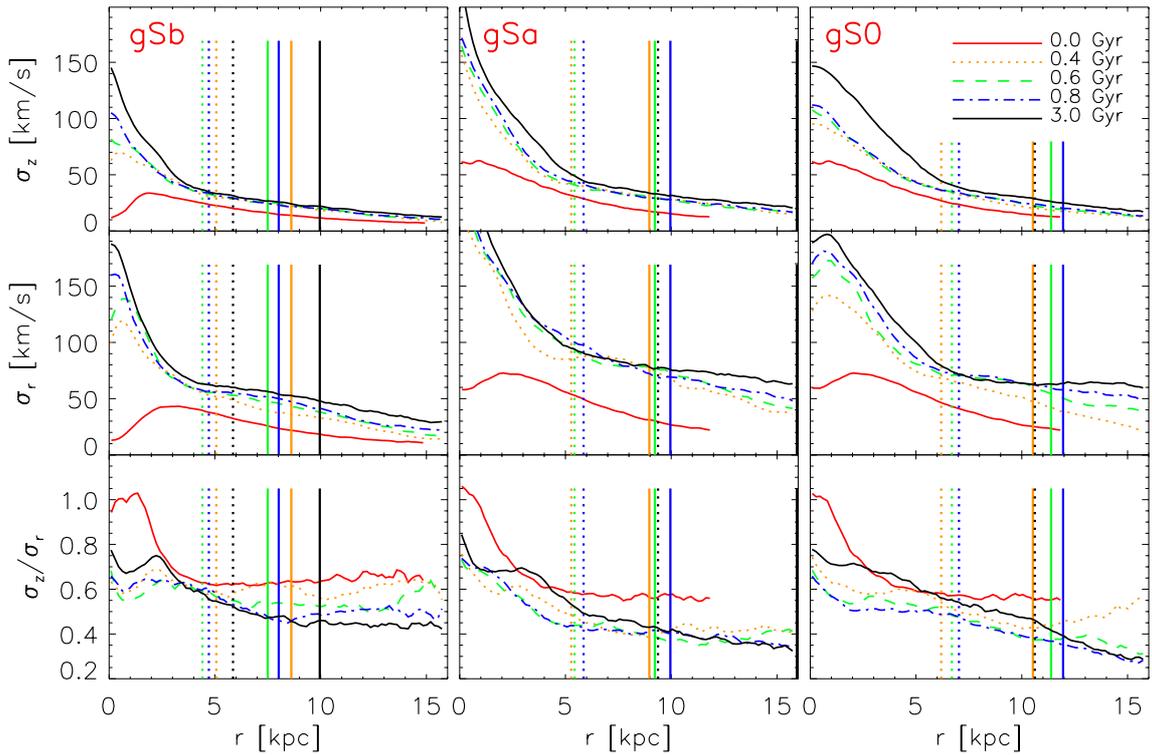}
\caption{
{\bf First row:} Time evolution of the radial vertical dispersion profiles, $\sigma_z(r)$, for the gSb, gSa, and gS0 models. Different colors and line styles correspond to the times indicated in the first row, rightmost panel. The bars' corotation (CR) and outer 2:1 Lindblad resonance (OLR) are shown by the dotted and solid vertical lines, respectively, with colors corresponding to the different times. {\bf Second row:} Same but for the radial velocity dispersion, $\sigma_r(r)$. {\bf Third row:} Same but for the ratio $\sigma_z/\sigma_r$. The slope in $\sigma_z/\sigma_r$ outside the bar shows that the radial profile flattens more with time than the vertical one, while its overall decrease means that the disc heats more radially than vertically. 
\label{fig:gSall_sig}
}     
\end{figure*}

\subsection{Cosmological re-simulations}

We concentrate on the controlled models just described for most of the analyses we are about to do, since these discs are easier to analyze given their constrained nature, lack of gas accretion and mergers. However, we also will use much more realistic simulations, following the self-consistent assembly of galactic discs in the cosmological context, in order to test the validity of our results. These numerical experiments were first presented by \cite{martig12}, where the authors studied the evolution of 33 simulated galaxies from $z=5$ to $z=0$ using the zoom-in technique described in \cite{martig09}. This technique consists of extracting merger and accretion histories (and geometry) for a given halo in a $\Lambda$-CDM cosmological simulation, and then re-simulating these histories at much higher resolution (150~pc spatial, and 10$^{4-5}$ M$_{\odot}$ mass resolution), replacing each halo by a realistic galaxy containing gas, stars and dark matter. We have chosen to examine three simulations, which we refer to as the C1, C2, and C3 models, with approximately flat rotation curves and circular velocities $V_c\sim200$, 180, and 200~km/s, respectively. The bulge/bar/disc decomposition and disc mass growth for the three galaxies are presented in the third row of Fig.~25 (C1), the first row of Fig.~29 (C2), and the bottom row of Fig.~27 (C3) by \cite{martig12}. 

These simulations differ in several ways from the gSb, gSa, and gS0 models described above: (i) the discs grow self-consistently as the result of cosmological gas accretion from filaments and gas-rich mergers, as well as merger debris, (ii) the simulation techniques are different both in the way gravitational forces are computed (Particle Mesh vs hierarchical tree) and in the way gas dynamics is treated (sticky-particles vs SPH), and (iii) the spiral and bar instabilities are present for a cosmological time, thus the effects we see are not the result of recent bar formation, for example.   

\begin{figure*}
\centering
\includegraphics[width=18cm]{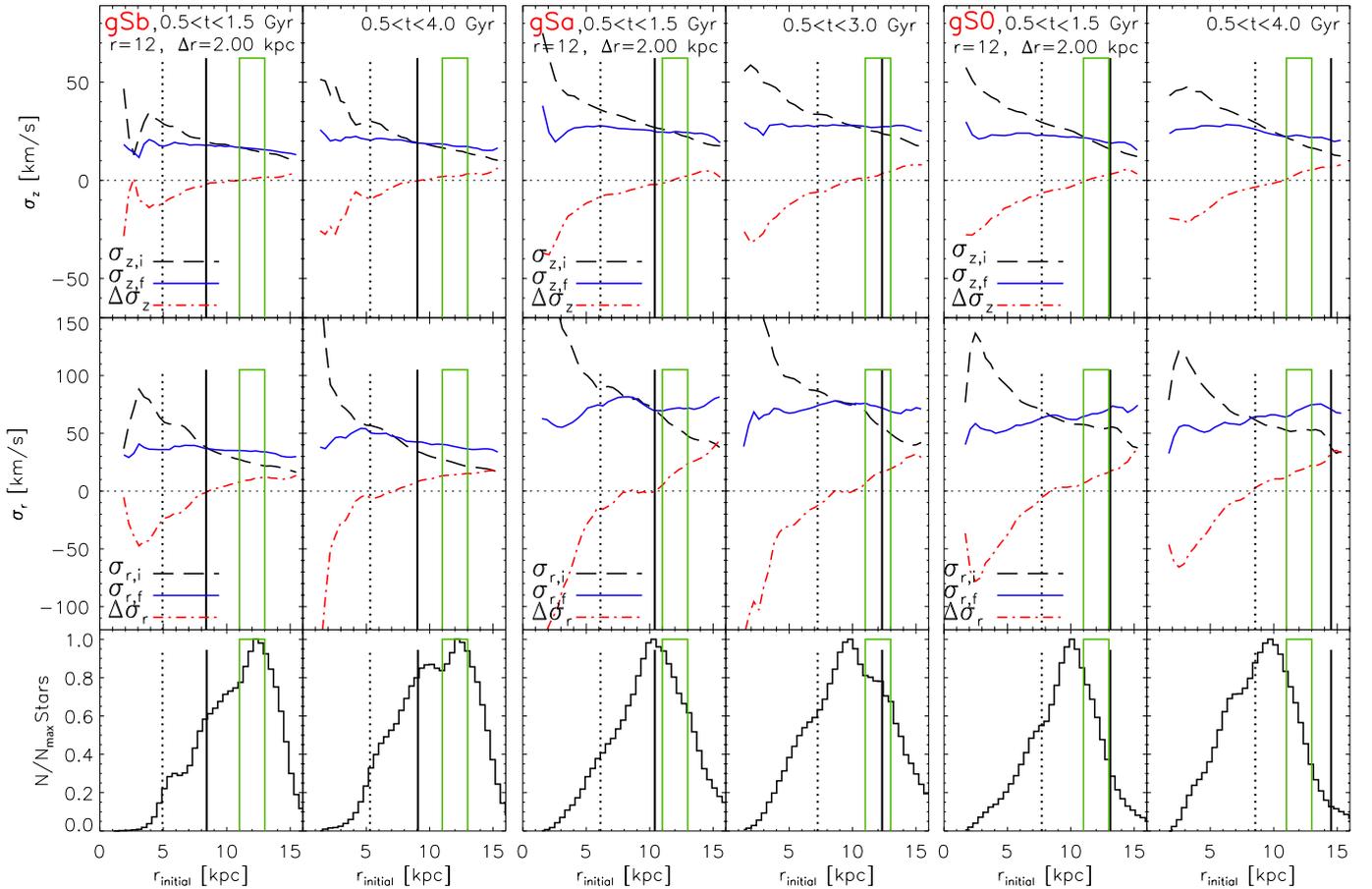}
\caption{
{\bf First row:} Initial (dashed black), final (solid blue), and net (dash-dotted red) vertical velocity dispersions, $\sigma_{z,i}$, $\sigma_{z,f}$, and $\Delta \sigma_z$, respectively, as functions of the {\it initial} radius, $r_0$, for particles ending up in the green bin at the final time indicated in each panel. Two time periods for each of the three models are shown, as indicated. Migrating stellar samples decrease or increase their velocity dispersion is such a way as to match the value at the destination. {\bf Second row:} Same as the first row, but for the radial velocity dispersion. {\bf Third row:} The corresponding initial radial distributions of stars ending up in the green bin at the final time indicated in each panel. The bar's CR and OLR, measured at the final times, are shown by the vertical dotted and solid black lines. Most migrators come from the inner disc, with a strong contribution from particles originating near the bar's CR.
}
\label{fig:all_sig}      
\end{figure*}

\section{Disc thickening and velocity dispersions}
\label{sec:multiple}

Up until Section~\ref{sec:cosmo} we only analyze our constrained, Tree-SPH N-body simulations described in Section~\ref{sec:nbody}. As we have shown in \cite{minchev12a}, all of these models exhibit multiple patterns, giving rise to strong radial migration. To show the strength of the discs' asymmetric components, in the first column of Fig.~\ref{fig:1} we plot the $m=2$ Fourier amplitudes, $A_2/A_0$, as a function of radius for each galaxy, where $A_0$ is the axisymmetric component and $m$ is the multiplicity of the pattern. Different color curves show the time evolution during the first Gyr. $A_2$ indicate the bi-symmetric structure in the disc. For example, for the gS0 model the bar is identifiable by the smooth curve in the inner disc, which at later times has a maximum at $\sim3$~kpc and drops almost to zero at $\sim8$~kpc. Deviations from zero seen beyond that radius are due to the spiral structure. Note that the spirals are strongest for the gSa model, related to the initial 10\% gas fraction in its disc. The gSb model is the one with parameters closest to the MW: bar size $\sim3-4$~kpc, $v_c\sim220$~km/s, and a small bulge. The circular velocities and bar pattern speeds for all three models can be found in Fig.~2 by \cite{minchev12a}.

To see how much discs thicken, in columns 2-3 of Fig.~\ref{fig:1} we plot the edge-on view for each disc and bulge (separately) for earlier (0.25~Gyr) and later (4~Gyr for gSb and gS0, and 3~Gyr for gSa) times of their evolution, as indicated in each panel. Significant thickening is seen for all models, especially for gSa and gS0 due to their larger bars. The bulges are also found to expand, not significantly so for the gSb. Although from these plots it is evident that the discs thicken, it is important to see how this is reflected in their scale-heights. These are shown in columns 4-5 at a radius 12~kpc ($\sim2.5$ disc scale-lengths), which, when scaled, is comparable to the solar distance from the Galactic center. The dotted red line shows single exponential fits. 

Surprisingly, we find that, despite the strong radial migration, the initial scale-heights, $h\approx0.3$~kpc, hardly double in the 3-4 Gyr of evolution for gSa and gS0, respectively, and increase by only $\sim$~50\% in 4~Gyr for gSb. 

The disc thickening we see in Fig.~\ref{fig:1} must be related to an increase in the vertical velocity dispersion. To see how our discs heat, in the first two rows of Fig.~\ref{fig:gSall_sig} we show the time evolution of the vertical, $\sigma_z(r)$, and radial, $\sigma_r(r)$, velocity dispersion profiles for the stellar populations of the gSb, gSa, and gS0 models in the first, second and third columns, respectively. Initial bulge components are not considered. In each panel the different color curves show different times, as indicated in the top left panel. The dotted and solid vertical lines indicate the radial positions of the bar's corotation (CR) and 2:1 outer Lindblad resonance (OLR) with colors matching the different times. 

An interesting observation is that the $\sigma_r$ profiles flatten at later times, especially for the gS0 and gSa models, i.e., the galaxies with the larger bars and stronger migration histories. This is reflected in the ratio $\sigma_z/\sigma_r$, shown in the bottom row, where the later curves (blue and black) develop a negative slope. Moreover, $\sigma_r$ increases more in the outer parts of the disc, compared to the region near the bar's CR. In fact, we observe radial cooling near the CR for both the gS0 and gSa models: the solid-black curve falls below the dashed-green and dash-dotted blue ones. The reason for this is revealed in the next Sections.

\section{Cooling/heating of stellar samples migrating outwards/inwards}
\label{sec:cooling}

Stellar samples in the inner galactic discs possess high velocity dispersions in all three components. It has been expected that by migrating radially outwards, stars retain their random energies and thus thicken the disc. If this were true, then how do we explain the lack of sufficient disc thickening even for the strong large-scale disc mixing seen in Fig.~\ref{fig:1}? To answer this question, we consider the samples of stars used to estimate the scale-heights in Fig.~\ref{fig:1}. In the first row of Fig.~\ref{fig:all_sig} we plot the initial (dashed black), final (solid blue), and net (dot-dashed red) vertical velocity dispersions, $\sigma_{z,i}$, $\sigma_{z,f}$, and $\Delta \sigma_z$, respectively, as functions of the {\it initial} radius, $r_0$, for particles ending up in the green bin ($11<r<13$~kpc) at the end of the given time interval. This is done for each galaxy in the time intervals $0.5<t<1.5$ and $0.5<t<3$~Gyr ($0.5<t<4$~Gyr for gS0 and gSb), as indicated in the figure. The second row shows similar plots but for the radial velocity dispersion. The distribution of initial radii of stars ending up in the green bins at the corresponding times are shown in the third row. 

From the first two rows it is immediately apparent that stellar samples arriving from the inner discs decrease their velocity dispersions by as much as 50\% when coming from the smallest radii, with the effect diminishing the smaller the radial distance traveled. Conversely, stars coming from radii exterior to the green bin increase their velocity dispersion. Therefore, on the average, newcomers change their random energies in such a way as to approximately match the non-migrating population at the radius at which they arrive. Thus, even though at later times (second, fourth and sixth columns) a larger fraction of migrators enters the annulus under consideration (as seen in the bottom row), this has no effect on the velocity dispersion increase. 

\begin{figure*}
\centering
\includegraphics[width=16cm]{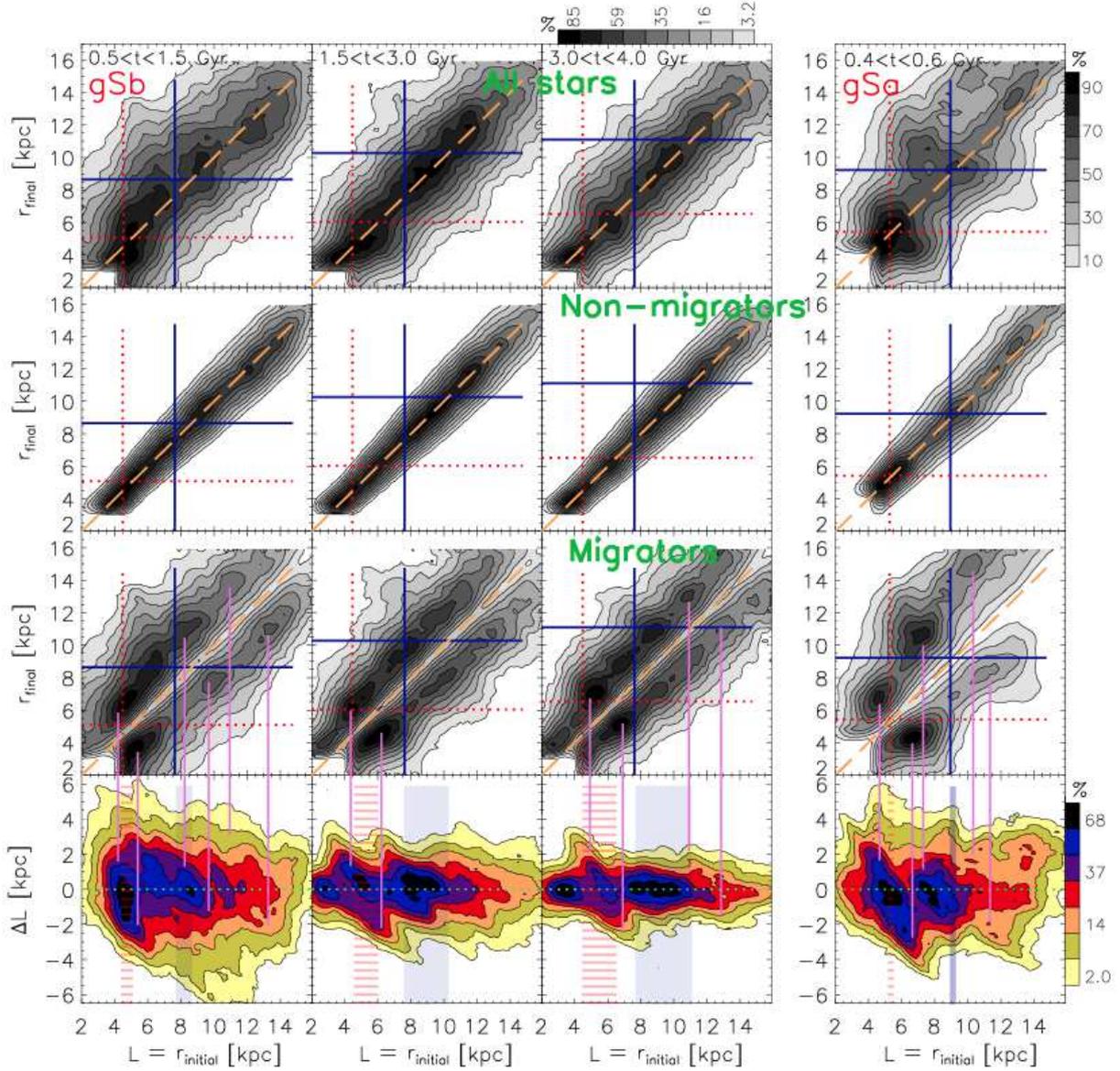}
\caption{
{\bf First row:} Contour density plots of the final versus initial radii for all particles in different time spans for the gSb or gSa models, as indicated. The orange dashed line shows the locus of non-migrating, cold particles. Nonuniform deviations from this line indicate that there exist preferential radii where migration efficiency is stronger, however, these are not clearly seen. {\bf Second row:} Non-migrating stars, extracted from the distributions above as described in the text. {\bf Third row:} The migrators only, obtained by subtracting the distributions in the second from those in the first row. The dotted-red and solid-blue vertical/horizontal lines in all panels indicate the positions of the bar's CR and OLR, respectively, at the initial/final times. Strong clumps can now be seen across the bar's CR for example. {\bf Fourth row:} Number density plots of the changes in angular momentum, $\Delta L$, versus the initial angular momentum (or equivalently initial radius), $L$, divided by the rotational velocity at each radius. The broken red and blue vertical strips indicate the location of the bar's CR and OLR, where their widths corresponding to the change during each time period.The pink lines connect some corresponding groups of migrated stars. The bar's CR is active throughout the entire simulations.
}
\label{fig:sig_map}      
\end{figure*}

The effect we just described is more dramatic for the radial velocity dispersion: for the gS0 (dissipationless) model the non-migrating stars (blue line portion inside the green bin) are hotter at the final time than the migrators, as evident by the positive slope of the solid-blue curve. Since the majority of stars arrive from the inner disc, the final annulus is being cooled, rather than heated by the migrators! We can see that, for all three models, stars arriving from inside the bar's CR are radially cooler than the non-migrators: note the positive slope in the blue curve inside the CR region (i.e., to the left of the dotted vertical line). Because for gS0 the radial bin considered happens to be between the CR and the OLR, we see the most cooling for this model. We find similar results when the bin is placed at a similar location with respect to the bar's resonances in the case of the gSa and gSb models (see Fig.~\ref{fig:frac}). 

The velocity distribution of migrating stars before and after they have migrated are smooth and appear approximately gaussian, as can be seen in Fig.~\ref{fig:vel_hist} in the Appendix.

\section{Separating migrators from non-migrators}
\label{sec:separating}

To assess the contribution of migrating stars to the intrinsic velocity dispersion of the disc we devise a simple procedure for separating migrators and non-migrators.
\newline
(1) We consider a disc annulus of a certain width at some initial and then final time. 
\newline
(2) We separate all stars into ``non-migrators" (those that are found in the selected radial bin at both the initial and final times) and ``migrators" (those that were not present in the bin initially but are there at the final time). 
\newline
(3) We do this for radial bins over the entire extent of the disc, overlapping every 0.3 kpc, for example, in order to get information at every radius.

In step (1) above we need to adopt a certain width for each annulus considered. We chose this as follows. Epicycle sizes are on the order of $a\sim\sigma_r/\kappa$, where $\kappa$ is the epicyclic frequency. In order to contain non-migrating stars, we allow for particles to oscillate around their guiding radii by $\pm a$, therefore, we select our bin size to be $2a$. Since both $\sigma_r$ and $\kappa$ are functions of radius, so is the value of $a$. In Fig.~\ref{fig:bins} we plot all three functions versus galactic radius. Note that the bin sizes increase with radius, where the effect is stronger for the hotter gSa and gS0.

We apply the above procedure to our gSb and gSa models and display the results in Fig.~\ref{fig:sig_map}. The first row shows contour density plots of the final versus initial radii for all particles in different time spans for the gSb or gSa models, as indicated. We exclude a small portion of the inner disc (white rectangle in the lower left corner) in order to display better the outer contours. The orange dashed line shows the locus of non-migrating, cold particles. Nonuniform deviations from this line indicate that there exist preferential radii where migration efficiency is stronger, however, these are not clearly seen. We note that the particle distributions presented here are not the true density, but a combination from all migrators and non-migrators resulting from the overlap of the radial bins we consider. However, deviation from the true density are small, as shown in Fig.~\ref{fig:ap1} and discussed in the Appendix. 

The second row of Fig.~\ref{fig:sig_map} presents the non-migrating stars extracted from the distributions shown in the first row as outlined in (1), (2), and (3) above. Deviation from the orange line are about 1-2~kpc, increasing outwards, in agreement with the bins-size variation shown in Fig.~\ref{fig:bins}.

The third row of Fig.~\ref{fig:sig_map} shows the migrators only, obtained by subtracting the distributions shown in the second row from those in the first row. The dotted-red and solid-blue vertical and horizontal lines indicate the location of the bar's CR and OLR at the initial (horizontal) and final (vertical) times. We can now see clearly overdensities on each side of the orange dashed line. We show how our results change by considering different bin size values in the Appendix (see Fig.~\ref{fig:ap2}).  

In the fourth row of Fig.~\ref{fig:sig_map} we plot number density contours of the changes in the specific angular momentum, $\Delta L$, versus the initial specific angular momentum, $L$, estimated during each time periods. Both axes are divided by the rotational velocity at each radius, therefore $L$ is approximately equal to the initial radius and $\Delta L$ gives the distance by which guiding radii change. The broken-red and solid-blue vertical strips indicate the location of the bar's CR and OLR, where their widths corresponding to the change during each time period related to the bar's slowing down. For all time periods considered we find a nice correlation with structure across the bar's CR, where stars inside the CR migrate outward and those outside it migrate inward. 
The pink lines connect some corresponding groups of migrated stars in the $r_{final}-r_{initial}$ and the $L-\Delta L$ planes. At the earlier times shown ($0.5<t<1.5$~Gyr, left column) the changes in angular momentum are the strongest due to the potent spirals and their interaction with the bar. At the intermediate times, as the spirals weaken, the typical feature of negative slope across the bar's CR is much better seen, in addition to some outer structure related to the spirals. Finally, at the later times ($3<t<4$~Gyr, third column) the bar's CR is clearly dominating. In general, although here we only show the entire simulation time for the gSb model and only the beginning of gSa, clumps across the bars' CR are invariably seen for all time periods and all three models. This continuous bar activity was shown also by \cite{minchev12a} (see their Fig.~7).

Contours are denser above the orange line in the $r_{final}-r_{initial}$ plane. This is most extreme during the early time period shown for the gSa model (rightmost column of Fig.~\ref{fig:sig_map}). The reason for this behavior is that, in the outer disc, stars are preferentially shifted outwards due to the decreasing stellar density. The effect is much stronger for gSa because the outward transfer of angular momentum for this simulation is much stronger than for the gSb disc, which \cite{minchev12a} related to the very efficient non-linear coupling among the bar and spiral waves of different multiplicity. Note that, since the total angular momentum of the disc must be conserved (with a small contribution to the halo and bulge), the outward transfer of angular moment needs to be balanced. This balance is achieved by transferring large amounts of mass inward of the bar's CR (with small negative changes in stellar guiding radii), in contrast to the exponentially decreasing density in the outer disc (but much larger increase in guiding radii). An indication of this process may be found in the strong clump of inward migrators just outside the bars' CR, seen in all plots but the latest time period for gSb; the reason for the latter is due to the saturation of migration efficiency in the outer disc in the absence of gas accretion (and thus spiral strength). 

\begin{figure}
\centering
\includegraphics[width=8.5cm]{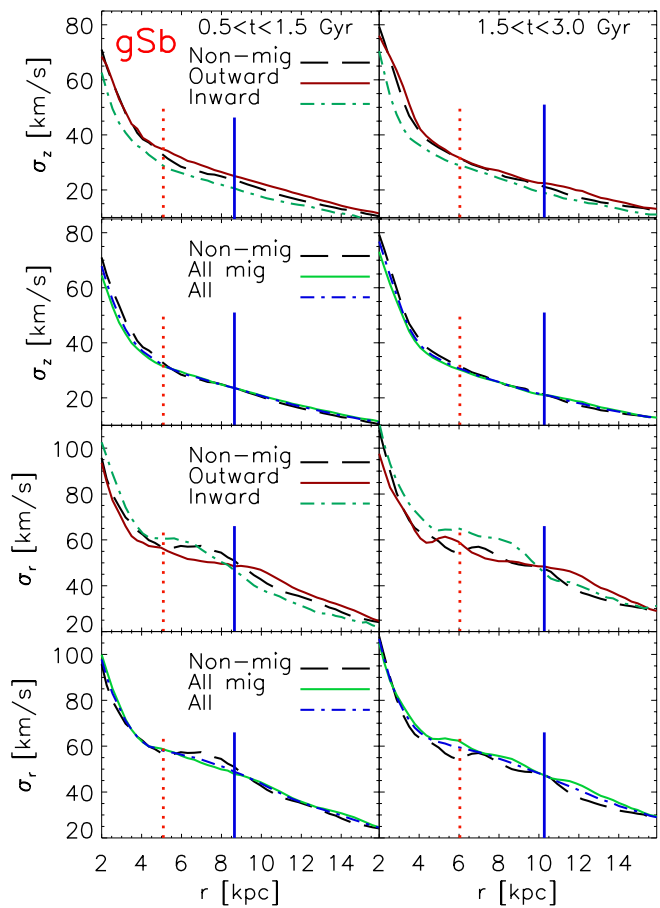}
\caption{
Migrators' contribution to the disc heating. {\bf First row:} radial profiles of the vertical velocity dispersion, $\sigma_z(r)$, for non-migrators (dashed black curve), outward going migrators (solid brown curve), and inward migrators (dash-dotted green curve) for the gSb model, estimated in the time periods indicated in the top row of each column. {\bf Second row:} $\sigma_z(r)$ estimated for non-migrators (dashed black curve), all migrators (i.e., inward and outward going, solid green curve), and the total population (dashed-dotted blue curve). {\bf Third and Fourth rows:} same as the first and second rows, respectively, but for the radial velocity dispersion profile $\sigma_r(r)$. An unexpected result is that outward migrators cool the disc inside the bar's OLR (and inward ones heat it), unlike in the case of $\sigma_z$. Some increase in $\sigma_z$ and $\sigma_r$ from migrators is seen beyond the OLR, while the inner disc is {\it cooled} both radially and vertically. Fig.~\ref{fig:frac} quantifies better the migrators' contribution to $\sigma_z$ and $\sigma_r$.
}
\label{fig:sig_mig}      
\end{figure}

\section{Migrators' contribution to the disc velocity dispersion}
\label{sec:contr}

By separating stars in the entire disc into ``migrators" and ``non-migrators", we are now in a position to investigate their individual effects on the disc velocity dispersion increase. We consider the time intervals $0.5<t<1.5$ and $1.5<t<3$~Gyr, corresponding to the first two columns in Fig.~\ref{fig:sig_map}. Because the heating in both radial and vertical directions mostly saturates around 1-2~Gyr (see Fig.~\ref{fig:gSall_sig}), the effect at later times is similar to what we find during the period $1.5<t<3$~Gyr. 

In the first row of Fig.~\ref{fig:sig_mig} we show the radial profile of the vertical velocity dispersion, $\sigma_z(r)$ for non-migrators (dashed black curve), outward going migrators (solid brown curve), and inward migrators (dash-dotted green curve) for the gSb model. The dotted-red and solid-blue vertical lines denote the bar's CR and OLR, respectively, at the final time of the time period considered. 

We find that inward migrators cool the disc and the outward ones heat it. This is consistent with the information contained in Fig.~\ref{fig:all_sig}, where we found that particles migrating to a final annulus at $r=12$~kpc arrive with $\sigma_z$ values slightly lower/higher than the non-migrating population (evident by the blue line's slightly negative slope in Fig.~\ref{fig:all_sig}). 

\begin{figure}
\centering
\includegraphics[width=8.5cm]{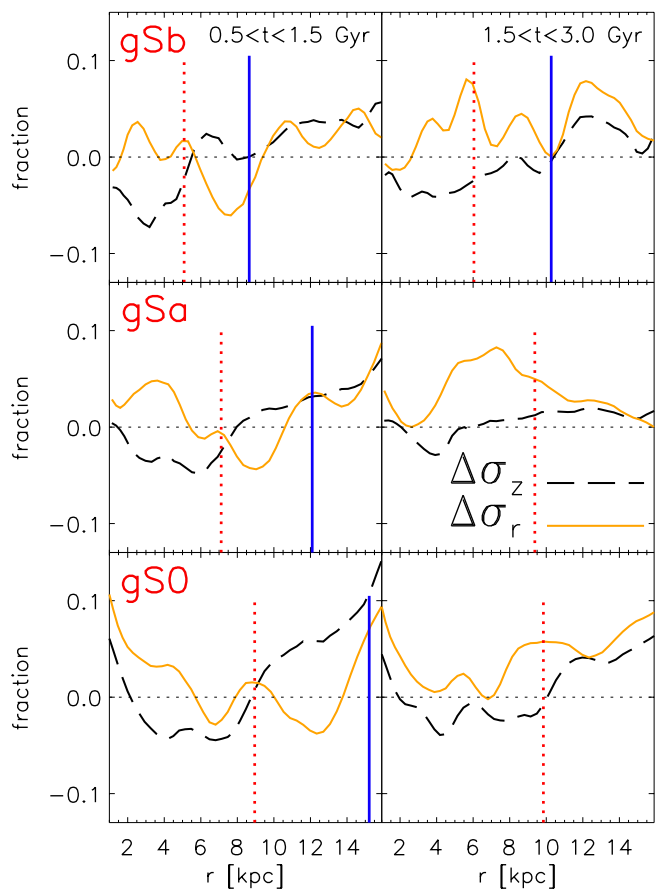}
\caption{
Fractional changes in $\sigma_z(r)$ and $\sigma_r(r)$ resulting from the migrated stars. The dashed black curve plots the quantity $\Delta\sigma_{\rm z}=(\sigma_{\rm z,all}-\sigma_{\rm z,non\_mig})/\sigma_{\rm z,all}$, where $\sigma_{\rm z,all}$ and $\sigma_{\rm z,non\_mig}$ are the vertical velocity dispersions obtained from the total population and the non-migrators, respectively. A similar expression estimates $\Delta\sigma_{\rm r}$, shown by the solid orange curve. The dotted-red and solid-blue vertical lines denote the bar's CR and OLR, corresponding to the final time shown in each time period. Note the inverse correlation between $\Delta\sigma_{\rm z}$ and $\Delta\sigma_{\rm r}$, indicating coupling between the radial and vertical motions of stars.
}
\label{fig:frac}      
\end{figure}

\begin{figure*}
\centering
\includegraphics[width=15cm]{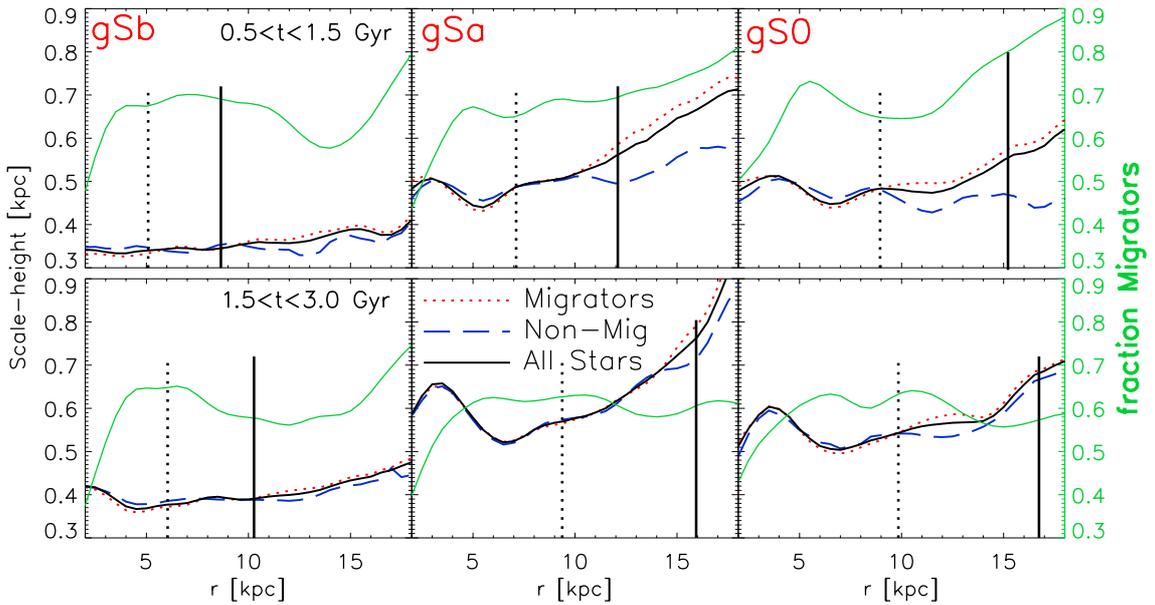}
\caption{
Scale-height radial profiles for migrators, non-migrators and all stars in the time intervals $0.5<t<1.5$~Gyr (top) and $1.5<t<3$~Gyr (bottom). The green solid curves indicate the fraction of migrators as functions of radius during each time period. The scale-height for the non-migrating stars is approximately constant with radius for all models (blue dashed curves, top). Due to the inner disc cooling and outer disc heating in the vertical direction from migrators (dotted-red curve, see also Fig.~\ref{fig:frac}), all discs flare (solid black curve).  
}
\label{fig:height}      
\end{figure*}

From the plots we just described (first row of Fig.~\ref{fig:sig_mig}) it is not clear how inward and outward migrators contribute to the velocity dispersion profile of the total population since we did not normalize to the number of particles in each group. How much $\sigma_z(r)$ of the total population is affected depends on the fraction of each kind of migrators at each radial bin. To demonstrate the true contribution of migrators, in the second row of Fig.~\ref{fig:sig_mig} we plot the vertical velocity dispersion for non-migrators (dashed black curve), all migrators (i.e., inward and outward going, solid green curve), and the total population (dashed-dotted blue curve). Interestingly, we find that the velocity dispersion from both types of migrators is very similar to that of the non-migrating particles, except for the inner parts of the disc for the earlier times shown. It is surprising that not only do migrators not heat most of the disc, but they even cool it inside the CR. Some increase in $\sigma_z$ can be seen outside the OLR.

We now examine the effect of migrators on the radial velocity dispersion profile, $\sigma_r(r)$. The third and fourth rows of Fig.~\ref{fig:sig_mig} are similar to the first two, except the plotted quantity is $\sigma_r(r)$. A distinct difference compared to the vertical velocity dispersion is the oscillatory behavior of $\sigma_r(r)$ found in all three curves in the third row. Comparing the third and fourth rows for the earlier time period, it becomes clear that the dominant effect inside the bar's OLR comes from the outward migrators, which here {\it cool, rather than heat the disc} as in the case of $\sigma_z(r)$. Just inside and outside the OLR, the effect of migrators is to reduce the wiggles seen in the non-migrators' $\sigma_r$, consequently resulting in a smooth radial velocity dispersion profile.      

To better quantify the contribution of migrators, we now calculate the fractional changes in $\sigma_z(r)$ and $\sigma_r(r)$ resulting from the migrated stars for the same time periods shown in Figs.~\ref{fig:sig_mig}. For the vertical velocity dispersion, we estimate these as 
\begin{equation}
\label{eq:1}
\Delta\sigma_{\rm z}=(\sigma_{\rm z,all}-\sigma_{\rm z,non\_mig})/\sigma_{\rm z,all}, 
\end{equation}
where $\sigma_{\rm z,all}$ and $\sigma_{\rm z,non\_mig}$ are the vertical velocity dispersions for the total population and the non-migrators, respectively. An equivalent expression estimates the contribution to the radial velocity dispersion. We plot the results in Fig.~\ref{fig:frac}. The dotted-red and solid-blue vertical lines denote the bar's CR and OLR, corresponding to the final time considered in each panel. It now becomes much clearer how exactly migrators affect the discs.

For all models shown $\Delta\sigma_{\rm z}$ (dashed black curve) shows oscillatory  behavior, changing abruptly from negative inside the CR, to positive outside it. The maximum deviations from zero for gSb, for example, are $\sim7\%$ cooling in the inner disc and $\sim4\%$ heating at larger radii, and appear consistent with Fig.~\ref{fig:sig_mig}. Clearly, such contribution to the vertical velocity dispersion would results in disc flaring, as we show in the next Section.

Now we look at the migrators' contribution to the radial velocity dispersion, $\Delta\sigma_{\rm r}$ (solid orange curve). Remarkably, there appears to be an inverse correlation between $\Delta\sigma_{\rm z}(r)$ and $\Delta\sigma_{\rm r}(r)$, where the two functions cross near the bars' CR and OLR. This oscillatory behavior with zeros near the CR and the OLR is present for all three models during the first time period shown, which is when the discs heat the most. Note that the crossing point near the OLR occurs above zero as the OLR is shifted more outward in the disc (top to bottom), due to the decreasing number of non-migrators and prevailing influence of outward migrators. Outside the OLR both $\Delta\sigma_{\rm z}$ and $\Delta\sigma_{\rm r}$ are positive, with a contribution of less then 5\% (gSb) or less than 10\% (gSa). At the later times the general trend of an inverse correlation for the radial and vertical velocity dispersions is still seen, although the contribution to the $\Delta\sigma_{\rm r}$ is larger.

The inverse functional behavior just described indicates that the vertical and horizontal motions of migrators are coupled, as expected for a system governed by non-linear dynamics.

\section{The effect on the discs' scale-height} 

We showed in Fig.~\ref{fig:frac} that the migrators' contribution to the vertical velocity dispersion increase is small and mostly in the outer disc. This suggests that our discs flare. To find out if this is really the case, we now measure the scale-height at different radial bins in the gS0, gSa and gSb discs.

In Fig.~\ref{fig:height} we show radial profiles of the discs' scale-height for migrators, non-migrators and all stars in the time intervals $0.5<t<1.5$~Gyr (top) and $1.5<t<3$~Gyr (bottom). As in the figures so far, we take the initial time at 0.5~Gyr (or after the bars have formed) in order to avoid the effect of the initial formation of asymmetric structure. The green solid curves indicate the fraction of migrators as functions of radius during each time period. Note that during the later time period ($1.5<t<3$~Gyr) the stars we refer to as ``non-migrators" may have experienced some migration previously. This is also true for the earlier times, i.e., some migration takes place before 0.5~Gyr, but the strongest effect occurs in the range $0.5<t<1.5$~Gyr. All our procedure ensures is that they stayed at the same radii during the later time interval.

Fig.~\ref{fig:height} shows that the scale-height for the non-migrating stars is approximately constant with radius for all models (blue dashed curves, top) during the earlier time period. In fact, for gS0 the scale-height for non-migrators shows a decrease with radius. However, due to the inner disc cooling and outer disc heating in the vertical direction from migrators (dotted-red curve, see also Fig.~\ref{fig:frac}), all discs flare (solid black curve). This is entirely consistent with Fig.~\ref{fig:frac}, where we found that the contribution of migrators to the vertical velocity dispersion is negative inside the bar's CR and positive in the outer disc.  

By comparing the variation of the migration fraction with radius (green solid curve) to the scale-height of the total population (black solid curve) we note an inverse correlation inside the bars' CR: the more migrators present at a given radius, the smaller the scale-height. This is especially obvious at the earlier time period and for the more efficient gSa and gS0. At larger radii, where all stars are in fact migrators (although fraction is smaller that 100\%, these stars have migrated out before t=0.5~Gyr), scale-height increases with migration fraction, which is expected. 

\begin{figure*}
\centering
\includegraphics[width=12.cm]{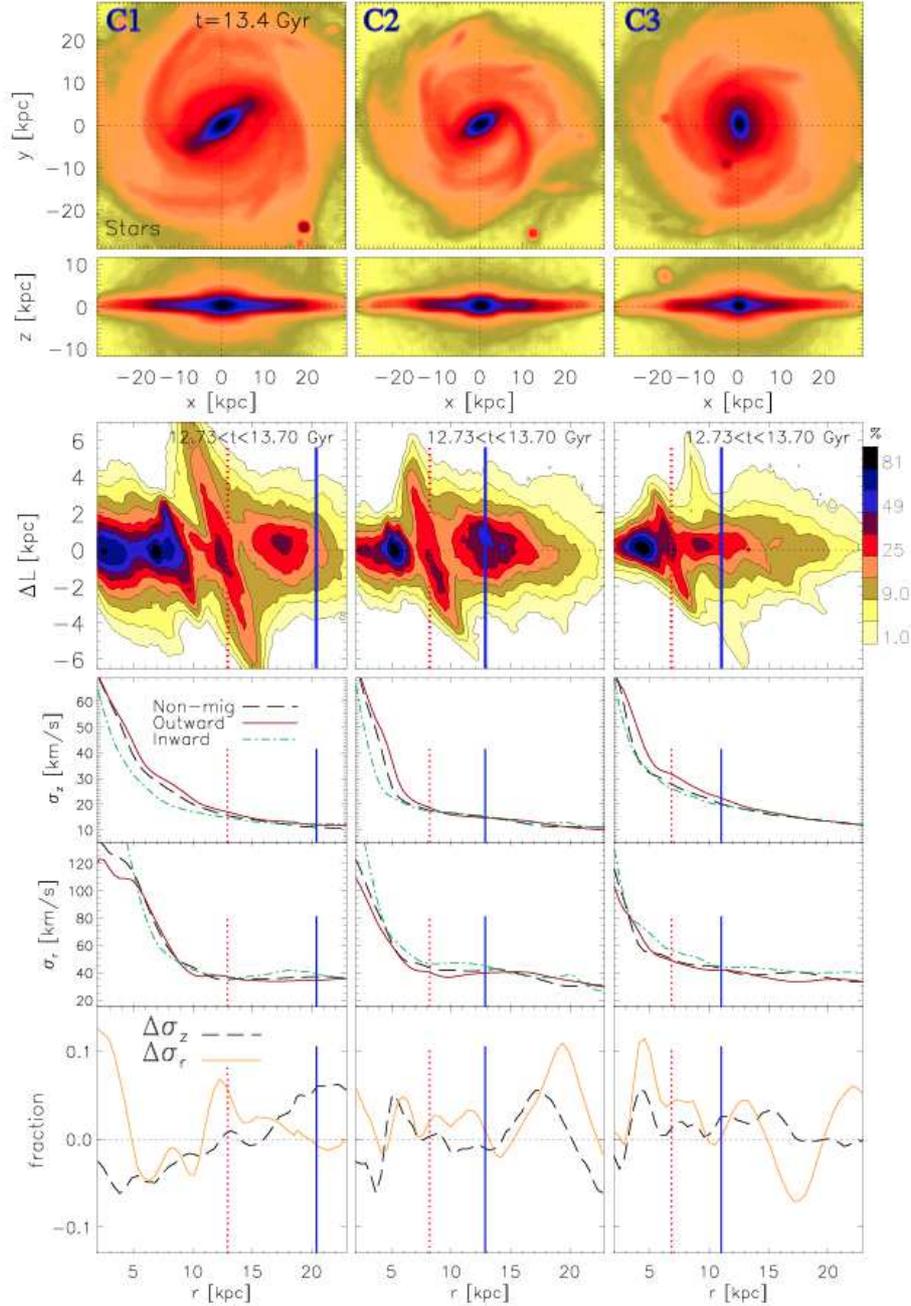}
\caption{
Application to three high-resolution cosmological re-simulations. {\bf First and second rows:} Face-on and edge-on density contours for the C1, C2, and C3 models. {\bf Third row:} Change in angular momentum as a function of disc radius: effective migration is seen for all models. {\bf Fourth row:} Vertical velocity dispersion profiles for non-migrators, outward and inward migrators. {\bf Fifth row:} Same but for the radial velocity dispersion. {\bf Bottom row:} Fractional contribution of migrators to the total vertical and radial velocity dispersions (as Fig.~\ref{fig:frac}). Results are very similar to the simpler models presented earlier. 
}
\label{fig:cosmo}      
\end{figure*}

\section{Application to cosmological re-simulations}
\label{sec:cosmo}

We now apply our newly developed technique for separating migrators and non-migrators to our three, much more realistic simulations in the cosmological context, described in Section~\ref{sec:sims}. To assess the effect of the internal perturbers and be able to compare to our previous results, here we only consider the last Gyr of evolution before $z=0$, thus, avoiding strong merger activity. We defer a more detailed study of these models to a future work.

During the $\sim1$~Gyr of evolution we study here, disc penetration by satellites occurs only at $(r,t)\approx(18$~kpc, 13.15~Gyr) for C1 and $(r,t)\approx($10~kpc, 13.4~Gyr$)$ for C3. There are no disc encounters within $r<30$~kpc for the C2 model. All of the aforementioned companions have mass ratios less than 1:100. Face-on and edge-on stellar density contours are plotted in the first and second rows of Fig.~\ref{fig:cosmo}, respectively, depicting the time $t=13.4$~Gyr. This is roughly in the middle of the timespan considered for the rest of the plots in the figure. Some small satellites are still seen in all three simulations. For C1 and C2 these are orbiting outside 20~kpc radius, but for C3 there is a small companion at $r\sim10$~kpc, located close to the plane at this time. 

\subsection{First results for migration in the cosmological context}

To see how much migration has taken place in the last Gyr of evolution, in the third row of Fig.~\ref{fig:cosmo}, for each model, we plot the changes in the specific angular momentum of stellar particles in the range $4<r<23$~kpc and $|z|\lesssim4$~kpc. The horizontal axis plots the initial angular momentum, $L_i$ (estimated at $t=12.73$~Gyr); this is divided by the circular velocity and, thus, approximately equals galactic radius. The vertical axis plots $\Delta L=L_f-L_i$, where, $L_f$ is the angular momentum estimated as $t=13.7$~Gyr. Stars at $\Delta L\sim4$, for example, gain angular momentum which brings them $\sim4$~kpc outward in the disc and conversely for negative $\Delta L$. The dotted-red and solid-blue vertical lines show the radial location of each bar's CR and OLR. The high peak near the OLR for the C3 model is caused by the satellite closest to the galactic center seen in the disc face-on view above. 

We note that this is the first time such plots are presented for simulations in the cosmological context. It is interesting that the strongest changes in angular momentum happen near the bar's CR, just as noted recently by \cite{minchev12a} (using the gSb, gSa, and gS0 models), despite the fact that here the bars have been formed for a long time (more than 6~Gyr) and are not transient. The strength of the bar correlates with the effect on the $r-\Delta L$ plane; the bar length is related to the pattern speed, as expected, with longer bars rotating slower, since bars exist within their CR radii. 

\subsection{Contribution of migrators to the increase in velocity dispersion}

Now that we have established that migration is efficient in these runs (in fact, very similar to our controlled simulations, see \citealt{minchev12a}, Fig.~7 and Fig.~\ref{fig:sig_map} in this paper), we can look for the effect of the migrating and non-migrating stars (during the time period considered) on the vertical and radial velocity dispersion profiles, just as we did in Figs.~\ref{fig:sig_mig} and \ref{fig:frac}. In the fourth row of Fig.~\ref{fig:cosmo} we plot the vertical velocity dispersion of inward (moving in) and outward (moving out) migrators, and the non-migrators, similarly to Fig.~\ref{fig:sig_mig}. Cooling from inward migrators and heating from outward ones is seen, mostly inside the bar's CR (dotted-red vertical). This is in agreement with what we found for our preassembled models, with the difference that the effect here is weaker in the outer disc, except for the C3 model, where a significant peak is seen between the CR and the OLR; the latter is most likely due to the small satellite mentioned earlier, taking place in the same radial range.

The fifth row of Fig.~\ref{fig:cosmo} is the equivalent of the top one, but for the radial velocity dispersion. As in our controlled models, cooling from outward migrators and heating from inward ones (the reverse of what happens for $\sigma_z$) occurs between the bar's CR and its OLR. Even in the case of C3, which is perturbed by the satellite at that radius, we see the same behavior. 

Finally, the bottom row of Fig.~\ref{fig:cosmo} plots the fractional contribution to the vertical and radial velocity dispersions from all migrators estimated from Eq.~\ref{eq:1}. Similarly to Fig.~\ref{fig:frac}, migrators contribute to $\Delta\sigma_z(r)$ by $\sim5\%$, cooling the disc inside the CR for C1, just as we found for the controlled models. For C2 and C3, the radial variation of $\Delta\sigma_z$ is more erratic, yet contribution is mostly less than $3\%$. The C1 model's dynamics is dominated by the presence of its large bar, which could explain its closer similarity to the controlled simulations. 

While outward migrators for all three models are radially cooler than the non-migrators in the region between the CR and the OLR (fifth row of Fig.~\ref{fig:cosmo}) just as the case of our constrained simulations, $\Delta\sigma_r(r)$ is only slightly below zero for C1 and C2 and slightly positive for C3. These differences are most likely due to tidal effects associated with the orbiting satellites seen in the face-on plots in the first row. An inverse relation between $\Delta\sigma_z(r)$ and $\Delta\sigma_r(r)$ for our cosmological simulations can still be seen. 

It is remarkable that both our isolated discs and the just described much more sophisticated  simulations (which also use a completely different simulation technique) exhibit very similar behavior. This makes our result, that migrators do not contribute significantly to the increase of the velocity dispersion of the disc, much stronger. We, therefore, suggest that our finding is a fundamental property of the secular evolution of galactic discs. It is possible that when stronger satellite perturbations are included, as is the case at the early stages of galaxy formation, the situation may be different. 

\section{Conservation of vertical and radial actions?}

Our findings so far suggest the existence of a conserved dynamical quantity for stars moving around the galactic disc, resulting in the conspiracy to approximately match (on the average) the kinematics expected at their destination. Since from Figs.~\ref{fig:all_sig},\ref{fig:sig_mig}, and \ref{fig:frac} it is clear that this is not the energy, the obvious candidates are the radial and vertical actions.

\subsection{The vertical action}
\label{sec:actions}

\cite{binney11} have shown that, to approximate how the vertical motion of disc stars is affected by their radial motion in an axisymmetric potential, the adiabatic approximation is valid, i.e. that the vertical action is adiabatically conserved in the epicyclic sense\footnote{This means that the vertical action calculated from the vertical potential $\Psi(z,t) = \Phi(R(t),z) - \Phi(R(t),z=0)$ is very nearly conserved if $R(t)$ is the horizontal motion calculated from the planar effective potential $\Phi(R,0)+L_z^2/R^2$.}. As first proposed by \cite{minchev11b} and recently confirmed by \cite{solway12}, the vertical actions of migrating stars are closely conserved, as well. 

Note that while in an axisymmetric disc the energy and angular momentum are conserved quantities (thus, conservation of actions is expected), since stars migrate due to resonant perturbation associated with disc asymmetries (such as the bar and spirals waves), the above mentioned classical integrals of motion are not necessarily applicable here. Nevertheless, following \cite{minchev11b}, we now show that even in the case of a simple galactic disc model, assuming conservation of vertical action is in good agreement with our results. 

We assume that for stars at a given radius, the vertical action is given by $J_z= E_z/\nu$, where $E_z$ and $\nu$ are the vertical energy and frequency, respectively. We also assume that this quantity is conceived (on average) as stellar samples shift guiding radii (i.e., migrate). This is a good approximation in the limit that (i) the vertical motion decouples from the planar motion (a good assumption when $z_{\max}\ll r$, which holds for most disc stars, but note that we did find evidence to the opposite in Sec.~\ref{sec:contr}), and (ii) the stars move outwards much more slowly than their vertical oscillations (not really the case). From Gauss' law and Poisson's equations, $\nu\sim\sqrt{2 \pi G \Sigma}$, therefore, as stars migrate radially, their vertical frequency changes as $\nu_{\rm mig}(r)\sim\exp(-r/2r_d)$, given that the stellar density varies as $\Sigma\sim\exp(-r/r_d)$. To conserve the mean vertical action of a set of stars migrating outwards, we then require that their mean vertical energy, $\langle E_{\rm z,mig}\rangle\sim\sigma_{\rm z,mig}^2$, decreases as $\nu$. Therefore, the vertical velocity dispersion of a stellar sample migrating radially outwards will decrease as $\sigma_{\rm z,mig}(r)\sim\exp(-r/4r_d)$, where $r$ is the radius at witch it ends up. 

On the other hand, the vertical velocity dispersion in the ambient (non-migrating) disc component changes with radius as $\sigma_{\rm z}(r)\sim\sqrt{\Sigma(r) h}$ (e.g., eq.~7 by \citealt{vanderkruit11}). Assuming the scale-height, $h$, remains constant, we find that $\sigma_z(r)\sim\exp(-r/2r_d)$. Thus, while a star's vertical energy does decay as it migrates outwards, it does not decay as fast as the underlying typical vertical energy of disc stars and, therefore, still increases the vertical temperature of the disc where it arrives. Similarly, stars migrating inwards would cool the disc. This is exactly what we see in Fig.~\ref{fig:sig_mig} at radii outside the bars' CR: the disc is heated/cooled by outward/inward migrators. We found that inside the CR the discs are, in fact, cooled vertically, since contribution is made only by inward migrators (see Figs.~\ref{fig:sig_mig} and \ref{fig:frac}). Similarly, in the disc outskirts outward migrators dominate, thus the vertical temperature is increased.

\subsection{The radial action}

For the simple galaxy disc model considered above, the same argument which lead us to conclude that the vertical action is conserved can also be applied in the plane. We assume a conservation of radial epicyclic action $J_p = E_p / \kappa$ where $E_p$ is the radial epicyclic energy and and $\kappa \sim 1/r$ is the radial epicyclic frequency, valid for a flat rotation curve. We, therefore, expect that orbits conserving $J_p$ lose energy $E_p$ at the rate of $1/r$.  

There are some differences when compared to the vertical motion analysis. We found that $E_z$ of migrators needs to change as $\nu_{\rm mig}(r)\sim\exp(-r/2r_d)$, which is the same as the expected drop in velocity dispersion. The radial epicyclic frequency, on the other hand, decreases as $1/r$. We may, therefore, expect that the changes in $E_z$ and $E_p$ for the same migrating sample are different, with migrators contributing more to the radial disc heating. 

At this point a deficiency in the simple model we consider is found. As revealed by Fig.~\ref{fig:sig_mig}, third row, in our simulations outward migrators radially cool the disc inside the bar's OLR instead of heating it, i.e., the opposite to the behavior of their vertical velocity dispersion. This effect is most prominent in the region between the CR and the OLR for all the models we study, while at $r\gtrsim r_{OLR}$ $\sigma_r(r)$ and $\sigma_z(r)$ behave similarly.

\subsection{The need for a more complex conserved quantity:\\ a function of both $J_z$ and $J_p$?}

We found that the conservation of epicyclic vertical action for migrators is consistent with our numerical results.  In the case of the horizontal motion, this simple model is less accurate, as we just discussed. This result is not really surprising, as (i) the adiabatic approximation is less adequate for horizontal motions than vertical ones even in the axisymmetric case \footnote{To improve this, \cite{binney11} changed the effective potential involved in defining the radial epicyclic energy to $\Phi(R,0)+(L_z+J_z)^2/R^2$.}, and (ii) it is also well known that the classical epicycle theory greatly underestimates the radial action even in the case of pure planar motion (see \citealt{dehnen99} for a better approximations of $J_p$).

Additionally, the adiabatic invariance of $J_p$ is not necessarily a good approximation for a time evolving system perturbed by multiple asymmetric patterns with stars migrating faster than their epicyclic oscillations. We note that the inverse correlation between $\Delta\sigma_z( r)$ and $\Delta\sigma_r( r)$ we found in Section~\ref{sec:contr} is suggestive of an exchange between the vertical and radial motions. An exchange between $J_z$ and $J_p$ can be achieved when vertical and horizontal resonances coincide in radius. Such locations may be found in the region inside the bar's CR and between the CR and the OLR, i.e., where we see the largest differences between $\Delta\sigma_z$ and $\Delta\sigma_r$. For example, \cite{combes90} have shown that the approximate coincidence in radius of the bar's vertical and horizontal 2:1 ILRs (inside CR) are responsible for the bar's peanut shape. We, therefore, speculate that a conserved quantity for migrating stars could be a function of both $J_z$ and $J_p$, perhaps in a similar fashion to the conservation of the sum $J_z + J_p$ proposed by \cite{sridhar96a,sridhar96b} in the context of an adiabatically growing galactic disc, which involves a time dependent perturbation as is the case of our discs perturbed by multiple patterns\footnote{Dubbed, "levitation", this mechanism is at work when the 2:2 resonance between the vertical and radial epicyclic frequencies moves through phase space, capturing along its path stars with large $J_p$ and small $J_z$ and releasing them into orbits with smaller $J_p$ and larger $J_z$ (and conversely).}.  More work is needed to find a possibly similar invariant for the more complicated case of migrating stars due to disc asymmetries.

\section{Discussion and conclusions}
 
In this work we have used Tree-SPH N-body (controlled) simulations and unconstrained, cosmological re-simulations, to study the effect of internal disc evolution on disc thickening. All analyzed models have been previously found, or here shown, to develop strong multiple patterns, resulting in efficient radial disc mixing. 

We devised a procedure for separating migrating from non-migrating stars in simulations. We found an excellent correspondence between the changes in angular momentum of stars as a function of initial radius and structure in the $r_{initial}-r_{final}$ space, thus, verifying that our separating technique is correct (see Fig.~\ref{fig:sig_map}). Using this simple procedure, we were able, for the first time, to study and contrast the kinematics and spatial distribution of migrators and non-migrators in galactic discs.

A number of recent works have assumed that stars migrating from the inner disc retain their high energies, thus populating a thick disc component. To test this expectation, we first compared the disc velocity dispersion profiles in the vertical and radial directions, $\sigma_z(r)$ and $\sigma_r(r)$, respectively, for the migrating and non-migrating populations. Surprisingly, we found that the migrators' contribution to the vertical velocity dispersion, $\Delta\sigma_{\rm z}$, is negative inside the bar's corotation (CR) radius, $r_{\rm cr}$, and less than 5\% in the range $r_{\rm cr}<r<3h_d$, where $h_d$ is the disc's scale-length (see Fig.~\ref{fig:frac}). This is related to the fact that migrating samples {\it change} their energy in such a was as to match closely the kinematical properties of the non-migrating population at the radius of arrival. We thus have shown that, not only do migrators not heat significantly, but they {\it cool} a large portion of the disc.  

Given the migrators' contribution to vertical heating in the outer discs and cooling at small radii, disc flaring could be expected. We showed that this is indeed the case, by plotting the disc scale-height  as a function of radius for migrators and non-migrators for each simulation. As anticipated, non-migrators exhibit an approximately constant scale-height in contrast to the flared disc composed of migrators (see Fig.~\ref{fig:height}). We have to note that this flaring is more significant outside the disc's initial radial distribution ($\sim12$~kpc), i.e., where all particles are migrators.  

Using a simple galactic disc model, where both the radial, $J_p=E_p/\kappa$, and vertical, $J_z=E_z/\nu$, actions, are conserved, we explained the decrease of velocity dispersion of outward migrators as disc cooling. In order to conserve the actions of a migrating sample of stars, their energies, and thus the velocity dispersions, need to change with radius similarly to $\kappa$ and $\nu$. While stellar samples migrating outwards lose vertical energy, $\langle E_z\rangle\sim\sigma_z^2$, their reduced $\sigma_z$ is still higher than what is expected from extrapolating $\sigma_z\sim\Sigma(r)$. This is consistent with Fig.~\ref{fig:sig_mig}, where we found that the outward migrators slightly heat the disc vertically and inward ones cool it. However, instead of a larger contribution from migrators to the radial velocity dispersion, as predicted by this simple model, we found that outward migrators {\it cool} the disc instead of heating it inside about three scale-lengths. This means that, in the approximation we considered, the radial energy decreases with radius faster than the epicyclic frequency, i.e., the cooling is even more drastic than predicted by the simple model from Section~\ref{sec:actions}. We suggested that a better conserved quantity is a function of both the vertical and radial actions.

We estimated that for the gSa and gS0 models, bulge contribution to the thick disc beyond $\sim$~2.5 disc scale-lengths is only about 1\%. For a MW-like bulge (the gSb model) no stars reach this radius. This is related to the bar's CR lying mostly outside such a small bulge (see Fig.~\ref{fig:1}). It should be kept in mind that here the initial distributions of the bulges were prescribed as Plummer spheres. As evident form the third row of Fig.~\ref{fig:sig_map}, a strong contribution comes from disc stars inside the bars' CR. 

Our controlled simulations did not consider gas accretion. In a more realistic case, the mass increase due to a continuous thin-disc formation would contract the thickened discs and further decrease the scale-heights \citep{bournaud09}, particularly in the outer discs, where for the gSb, gSa, and gS0 models all stars are migrators. To check this, we applied our technique to much more realistic, high-resolution, unconstrained simulations in the cosmological context, where discs build up self-consistently. Interestingly, we found very similar results to those obtained with our simpler models, where the migrators' contribution to the total increase in random motions is typically less then 3\%. 

We do not expect insufficient resolution to be affecting our conclusions for the reason that low resolution should result in more heating, not less, as we find here. Hence, increasing the (intermediate) resolution of our controlled simulations would give rise to even less disc thickening\footnote{Note that if the resolution is too high, then the simulations will have unrealistic dynamics -- real galaxies have sources of noise, such as molecular clouds, star clusters, mergers.}. Moreover, the recent work by \cite{roskar11} has shown that increasing the softening length from 50 to 500~pc does not result in any significant change in the disc dynamics, therefore, we expect that the migration efficiency is not affected. Finally, our results do not change significantly when we use the higher resolution cosmological re-simulations studied in Sec.~\ref{sec:cosmo}. In any case, we are not interested here so much in the overall heating, but in the difference between the increase of velocity dispersion of migrators and non-migrators. 

It is remarkable that both our isolated discs and the much more sophisticated simulations in the cosmological context (which also use a completely different simulation technique) exhibit very similar behavior. We believe the inability of migration to thicken discs is a fundamental property of secular disc evolution, irrespective of the migration mechanism at work. We note, however, that when mergers are involved (i.e., when the evolution becomes non-secular), the approximate conservation of the vertical and radial actions we find here, may not be valid anymore. We defer a study of this important topic to a future work. 

\subsection{Expected signatures of migration in galactic discs}

Considering the results of this work, we now outline three possible observables for the effects of radial migration. 

(i)  {\it Old disc flaring:}
The effect of migration, driven by internal disc evolution, on the disc heating is mostly seen in the outer discs. The reason for this is the following. As we found in Sec.~\ref{sec:cooling}, as stars migrate outwards their velocity dispersion decreases as to approximately match the energy of stars at the final radius (which did not migrate). However, a small amount of energy may be retained, with the effect being stronger, the farther the stars come from. In other words, while stars do arrive with slightly larger energy at an outer radius, the effect is negligible, unless the change in guiding radius is very large. It thus appears that migration would always work as to increase/decrease the scale-height in the outer/inner discs, giving rise to flaring. We showed in Section~\ref{sec:cosmo} that the situation may be different when self-consistently growing discs are considered (see Fig.~\ref{fig:cosmo}). In any case, we do expect that older stellar populations, being exposed to migration mechanisms the longest, would exhibit flared vertical profiles, even if the total mass in the disc would not flare. Since this effect is related to the conservation of the average vertical action of stars, in the case of an active merger history, this signature may have been erased.    

(ii) {\it Flattening of the average vertical action radial profile:}
As pointed out by \cite{minchev11b}, an important consequence of the conservation of actions for migrating stars is that radial migration also mixes the radial distributions of the actions $\langle J_p\rangle$ and $\langle J_z\rangle$: the more the disc mixes, the weaker their variation with radius. This may offer a way to constrain the migration history in the Milky Way. Estimates of the average (better conserved) vertical action in observations for different populations of stars should reveal different variation with radius (flattening) for older groups of stars.  

(iii)  {\it Structure in vertical chemical gradients:}
While we here showed that a {\it kinematically} thick disc is not expected to result from radial migration, signatures of migration may be found in the current chemical distribution of stars, since outward migrators are preferentially deposited at high altitudes, the reverse being true for inward ones (see Fig.~\ref{fig:sig_mig}). Thus, for a given chemical evolution model for the Galaxy, predictions can be made for vertical chemical gradients as a function of radius as observed today.

\acknowledgements

We would like to thank P. Di Matteo, F. Combes, D. Pfenniger, L. Athanassoula, C. Chiappini,  R. de Jong, and M. Steinmetz for helpful discussions. We also thank the anonymous referee for valuable suggestions which have greatly improved the manuscript.


\section{Appendix}

\subsection{Velocity distributions of migrators and non-migrators}

\begin{figure}
\centering
\includegraphics[width=8.5cm]{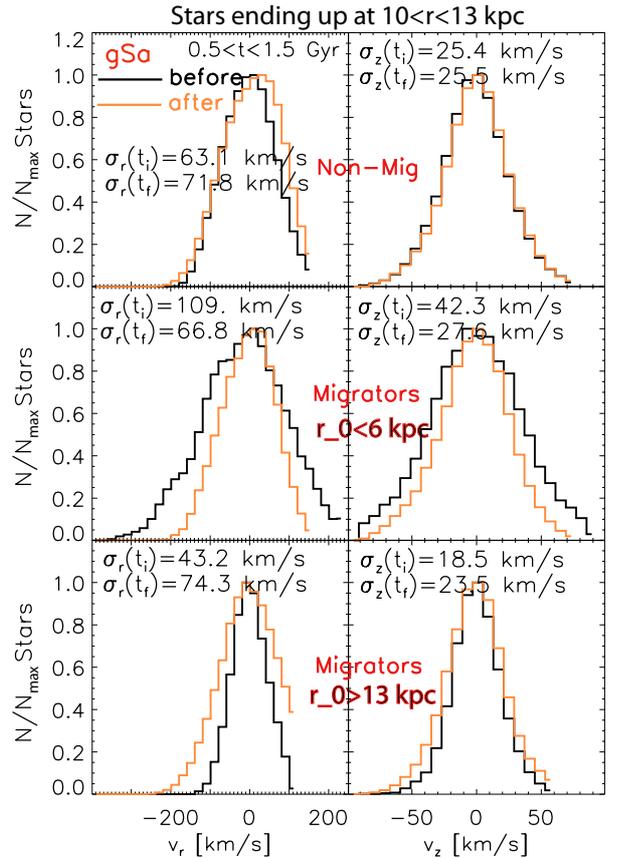}
\caption{
Distributions of migrators and non-migrators ending up in a radial bin in the range $10<r<13$~kpc, estimated in the time period $0.5<t<1.5$~Gyr for the gSa model. {\bf First row:} The radial (left) and vertical (right) velocity distributions for stars which stay in the radial bin during the entire time. An increase of velocity dispersion is seen with time. {\bf Second row:} Stars which came from $r<6$~kpc. Note that their velocity dispersions drop by about 40\%. {\bf Third row:} Stars originting from $r>13$~kpc. These are found to heat up upon arriving at the final radius. In all cases, final distributions (orange histogram) roughly match for each velocity component.   
}
\label{fig:vel_hist}      
\end{figure}

\begin{figure*}
\centering
\includegraphics[width=15cm]{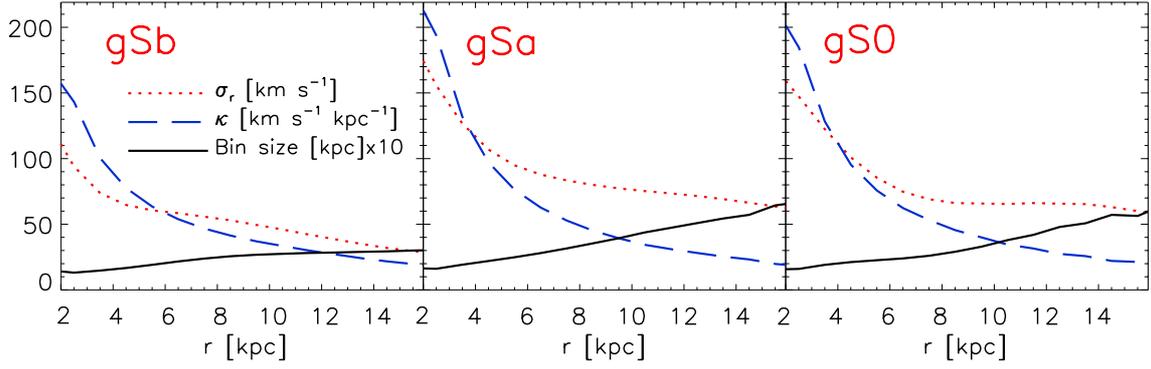}
\caption{
Variation of radial bins size with radius estimated as $2\sigma_r/\kappa$ (black curve in units of 10~km/s), where $\sigma_r$ is the radial velocity dispersion and $\kappa$ is the epicyclic frequency. The dotted red and solid blue curves show $\sigma_r$ and $\kappa$, respectively. Note that the bin sizes increase with radius.
}
\label{fig:bins}      
\end{figure*}

In Fig.~\ref{fig:all_sig} we showed that as stellar samples migrate, they decrease/increase their velocity dispersions both in the radial and vertical directions when coming from the inner/outer disc, respectively, so as to approximately match the kinematics of non-migrating stars in the final radial bin. Fig.~\ref{fig:vel_hist} offers a different view of this process. We consider the gSa model in the time interval $0.5<t<1.5$~Gyr (same as the third column of Fig.~\ref{fig:all_sig}). The first row of Fig.~\ref{fig:vel_hist} shows the radial (left) and vertical (right) velocity distributions for stars which stay in the radial bin during the entire time. An increase of velocity dispersion is seen with time. The second row shows stars which came from $r<6$~kpc. Note that their velocity dispersions drop by about 40\%. The third row plots stars originating from $r>13$~kpc. These are found to heat up upon arriving at the final radius. In all cases, final distributions (orange histogram) roughly match for each velocity component.

\subsection{Particle density from selection procedure}

\begin{figure*}
\includegraphics[width=18cm]{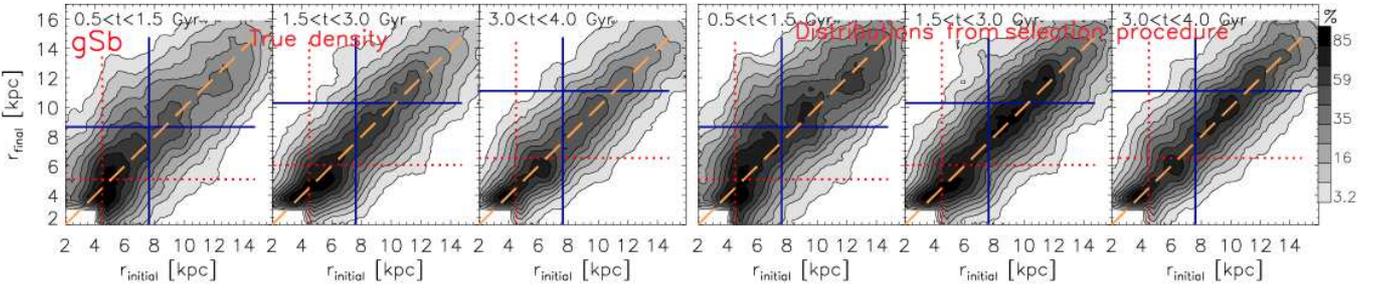}
\caption{
Contrasting the true density of the gSb model (panels 1-3), estimated directly from the initial and final radii of stars in the disc for the times shown, and the density obtained in the method described in Section~\ref{sec:separating} (panels 4-6, same as top row of Fig.~\ref{fig:sig_map}).
}
\label{fig:ap1}      
\end{figure*}

Here we discuss in more detail our procedure for separating migrators and non-migrators. Fig.~\ref{fig:ap1} contrasts the true density (panels 1-3), estimated directly from the initial and final radii of stars in the disc for the times shown, and the density obtained in the method described in Section~\ref{sec:separating} (panels 4-6, same as top row of Fig.~\ref{fig:sig_map}). Note that the overall shape of contours remains the same despite the overlap of particles resulting from our selection method. Considering that the contour levels are the same, we see that the true density drops faster with radius. This is due to the fact that the overlap of stars involves more stars coming from the inner disc, due to the asymmetric drift. It is clear that this overlap will not affect velocity measurements since contribution from multiple stars would cancel out. In any case, we remove multiples prior to estimating the velocity dispersions and disc scale-height. 

\subsection{Radial bin overlapping}

To achieve smooth variation with radius when estimating velocity dispersion profiles, etc., for all figures presented in the paper we overlapped bins every 0.3~kpc. Note that a typical radial bin width is about 2-3~kpc, increasing with radius (see Fig.~\ref{fig:bins}). In Fig.~\ref{fig:ap2} we examine the effect of changing the distance between the radial bins. The top, right pair of panels is the same as the top row of Fig.~\ref{fig:frac} (bin overlap of 0.3~kpc), showing the fractional changes in the vertical and radial velocity dispersions due to migrators, $\Delta\sigma_{\rm z}(r)$ and $\Delta\sigma_{\rm r}(r)$, respectively. We contrast four possibilities: bin overlap of 0.1, 0.3, 0.5 and 1~kpc. The overall shapes of the functions plotted remain the same, while, naturally, structure is lost with an increase in the distance between overlapping bins, i.e., for 0.1 to 1~kpc.

\subsection{The effect of different bin sizes}

To construct the figures in this paper we have estimated the bin size as $2a$, where $a\sim\sigma_r/\kappa$ is the approximate epicyclic amplitude at a given radius, with $\sigma_r$ and $\kappa$ the radial velocity dispersion and epicyclic frequency, respectively. In order to contain non-migrating stars, we allow for particles to oscillate around their guiding radii by $\pm a$, therefore, we select our bin size to be $2a$. Since both $\sigma_r$ and $\kappa$ are functions of radius, so is the value of $a$. 

In Fig.~\ref{fig:ap2} we show how the migrators' contribution to the disc velocity dispersion (as shown in Fig.~\ref{fig:frac}) changes with bin size. In addition to the choice of $2a$ used in Fig.~\ref{fig:frac} (middle two columns here), in the first two rows we also consider a bin size of $a$ (columns 1-2) and $3a$ (columns 5-6) for the gSb and gSa models and earlier and later times, as indicated. The bottom row shows the effect of a constant bin size of 1, 2, and 3~kpc for the gSb model. 

In all cases, the innermost disc appears to be affected the most by the change in bin size. Concentrating on the earlier times (when most of the velocity dispersion increase takes place), we note that the oscillatory inverse functional behavior of $\Delta\sigma_{\rm z}(r)$ and $\Delta\sigma_{\rm r}(r)$ is seen for all bin sizes.

\begin{figure*}
\includegraphics[width=16cm]{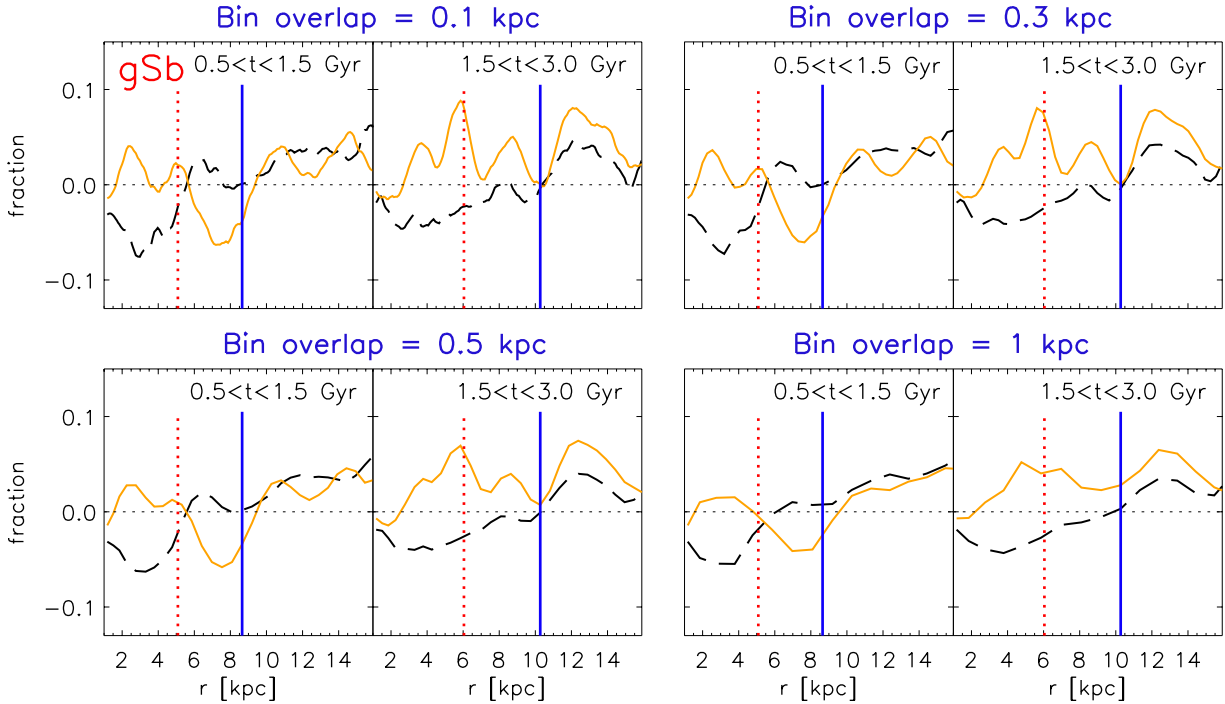}
\caption{
The effect of different bin overlapping used in the selection procedure. The top, right pair of panels is the same as the top row of Fig.~\ref{fig:frac} (bin overlap of 0.3~kpc), showing the fractional changes in the vertical and radial velocity dispersions due to migrators, $\Delta\sigma_{\rm z(r)}$ and $\Delta\sigma_{\rm r(r)}$, respectively. We contrast four possibilities: bin overlap of 0.1, 0.3, 0.5 and 1~kpc. The overall shapes of the functions plotted remain the same, while, naturally, structure is lost with an increase in the distance between overlapping bins, i.e., for 0.1 to 1~kpc. 
}
\label{fig:ap2}      
\end{figure*}

\begin{figure*}
\includegraphics[width=18cm]{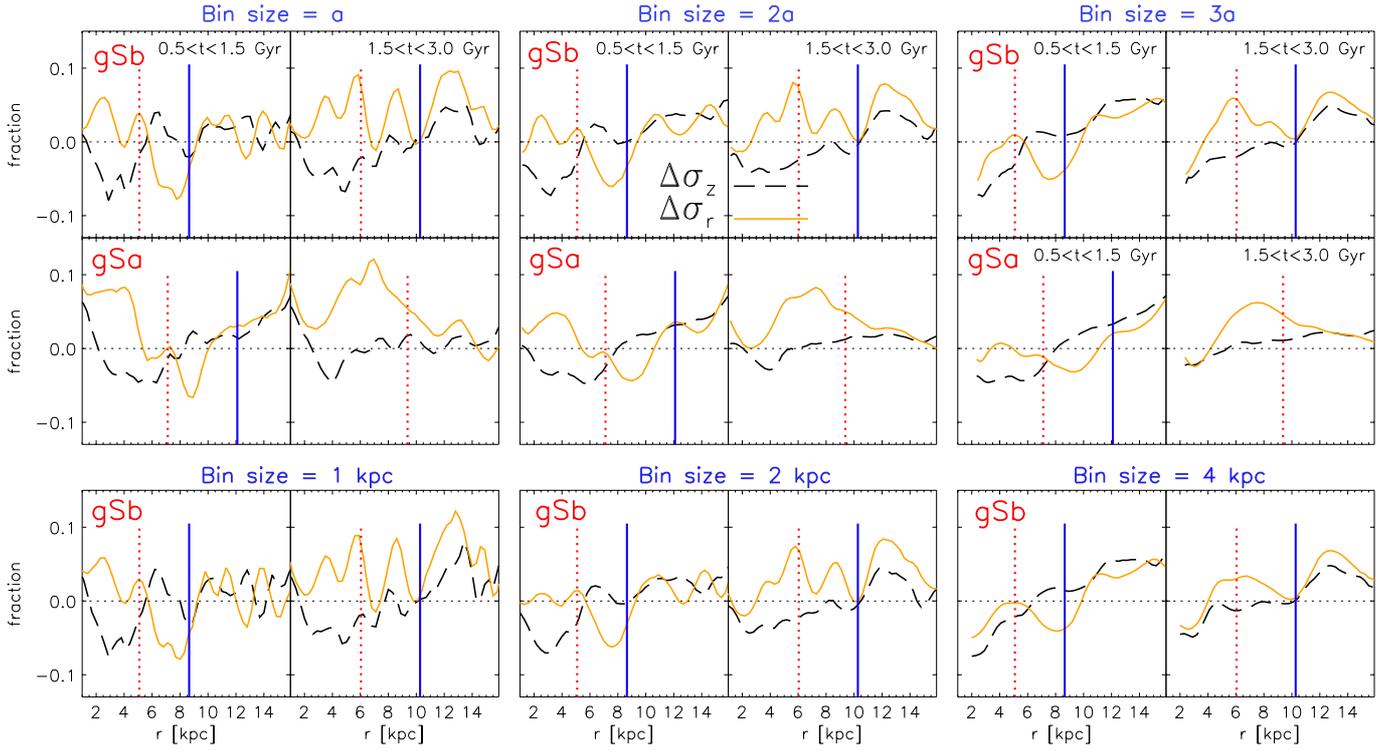}
\caption{
Changes in Fig.~\ref{fig:frac} due to the use of different bin sizes in the selection procedure. In addition to our standard choice of $2a$ (middle two columns here), in the first two rows we also consider a bin size of $a$ (columns 1-2) and $3a$ (columns 5-6) for the gSb and gSa models and earlier and later times, as indicated. The bottom row shows the effect of a constant bin size of 1, 2, and 3~kpc for the gSb model. The overall behavior is preserved in all cases.
}
\label{fig:ap2}      
\end{figure*}

\end{document}